\documentclass[journal,a4paper,10pt]{IEEEtran}
\usepackage{amsmath,amssymb,amsfonts}
\usepackage{graphicx}                  
\usepackage{url}                 
\usepackage{fontenc}
\usepackage{multirow} 
\usepackage{theorem}
\usepackage{array}

\usepackage{color,cite}
\usepackage{algorithmic}
\usepackage{algorithm}
\usepackage{epstopdf}
\usepackage{lipsum}
\usepackage{esint}
\usepackage{mathrsfs}
\usepackage[english]{babel}
\usepackage[table,xcdraw]{xcolor}
\usepackage{booktabs}
\usepackage{multirow,makecell}
\usepackage{longtable}
\usepackage{relsize}
\usepackage{comment}

\usepackage{colortbl}

\definecolor{Gray}{gray}{0.75}
\definecolor{Blue}{rgb}{0 ,0.99,0.98}
\definecolor{LightCyan}{rgb}{0.88,1,1}
\newcolumntype{b}{>{\columncolor{Gray}}c}
\newcolumntype{a}{>{\columncolor{Blue}}c}

\usepackage{array}
\newcolumntype{P}[1]{>{\centering\arraybackslash}p{#1}}
\newcolumntype{M}[1]{>{\centering\arraybackslash}m{#1}}

\begin{document}

\title{Blockchain and 6G: The Future of Secure and Ubiquitous Communication}
\author{{Ali Hussain Khan, Naveed UL Hassan, Chau Yuen, Jun Zhao, Dusit Niyato, Yan Zhang and H. Vincent Poor}
\thanks{A. H. Khan and N. U. Hassan are with the Department of Electrical Engineering at Lahore University of Management Sciences (LUMS), Lahore, Pakistan 54792. (Emails: 18060048@lums.edu.pk, naveed.hassan@lums.edu.pk).}
\thanks{C. Yuen is with the Engineering Product Development at the Singapore University of Technology and Design (SUTD), 8 Somapah Road, Singapore 487372. (Email: yuenchau@sutd.edu.sg).}
\thanks{J. Zhao and D. Niyato are with the School of Computer Science and Engineering at Nanyang Technological University, Singapore 639798. (Emails: junzhao@ntu.edu.sg, dniyato@ntu.edu.sg).}
\thanks{Y. Zhang is with the Department of Informatics at University of Oslo, Oslo, Norway 0315. (Email: yanzhang@ifi.uio.no).}
\thanks{H. V. Poor is with the Department of Electrical and Computer Engineering at Princeton University, NJ 08544. (Email:poor@princeton.edu).}

\thanks{This work was supported in part by LUMS Faculty Initiative Fund (FIF), in part by the U.S. National Science Foundation under Grants CCF-1908308 and ECCS-2039716, in part by WASP/NTU grant M4082187 (4080) and Singapore Ministry of Education (MOE) Tier 1 (RG16/20), and in part by A*STAR under its RIE2020 Advanced Manufacturing and Engineering (AME) Industry Alignment Fund – Pre Positioning (IAF-PP) (Grant No. A19D6a0053). Any opinions, findings and conclusions or recommendations expressed in this material are those of the author(s) and do not reflect the views of A*STAR.}

\thanks{Copyright (c) 20xx IEEE. Personal use of this material is permitted. However, permission to use this material for any other purposes must be obtained from the IEEE by sending a request to pubs-permissions@ieee.org.}
}

\markboth{IEEE WIRELESS COMMUNICATIONS, VOL. XX, NO. X}%
{Shell \MakeLowercase{\textit{et al.}}: Bare Demo of IEEEtran.cls for IEEE Journals}

\maketitle

\begin{abstract}
The future communication will be characterized by ubiquitous connectivity and security. These features will be essential requirements for the efficient functioning of the futuristic applications. In this paper, in order to highlight the impact of blockchain and 6G on the future communication systems, we categorize these application requirements into two broad groups. In the first category, called Requirement Group I \mbox{(RG-I)}, we include the performance-related needs on data rates, latency, reliability and massive connectivity, while in the second category, called Requirement Group II \mbox{(RG-II)}, we include the security-related needs on data integrity, non-repudiability, and auditability. With blockchain and 6G, the network decentralization and resource sharing would minimize resource under-utilization thereby facilitating RG-I targets. Furthermore, through appropriate selection of blockchain type and consensus algorithms, RG-II needs of 6G applications can also be readily addressed. Through this study, the combination of blockchain and 6G emerges as an elegant solution for secure and ubiquitous future communication.

\end{abstract}

\section{Introduction}

As 5G is approaching commercial readiness, 6G vision papers have started to appear in the literature \cite{saad2019vision,tariq2019speculative,giordani20196g,dang2019human,xiaohutowards}. These papers identify some key 6G applications and services such as, Human Bond Communication (HBC), Multi-sensory eXtended Reality Applications (XR), Wearable Technology based Futuristic Applications (WTech), Large-scale connected autonomous systems (LS-CAS), and greater support for several vertical domains. These applications have very stringent requirements of data rate, latency and reliability. The nature of data collected by several 6G applications will be increasingly sensitive and critical. The successful adoption of 6G applications by the users would therefore require strict data security guarantees. Blockchain is a distributed ledger technology where cryptography and hash functions are used to form a chain of data blocks, created when an event occurs and  verified in a decentralized way using consensus algorithms \cite{sadek2020blockchain}. Blockchain, initially only used in cryptocurrencies is now being used in other application domains like smart grid, connected vehicles, and Internet of Things. \cite{hassan2019blockchain,kang2019toward,daiblock2019,reviewer_ref}.

\indent Blockchain is believed to be a key technology in the 6G applications \cite{tariq2019speculative, saad2019vision, dang2019human}. The stringent network performance requirements of these applications will require support of technologies like Reconfigurable Intelligent Surafaces (RIS), TeraHertz (THz) communication, Artificial Intelligence (AI) and small cell networks. To enable efficient combination of these technologies for the provisions of resources to achieve the performance requirements, collaboration and coordination in a transparent and trustless environment is needed. These technologies also require dense network deployments which will lead to more infrastructure and complicated network deployment. Network decentralization will be needed to simplify the network deployment. Blockchain will provide the desired transparency and trustlessness in the decentralized network. Blockchain will also provide the strict security requirements of the future communication systems because of its in-built security features.  

\indent Based on the application requirements, the decentralization, security and scalability of blockchain can be fine tuned by selection of appropriate blockchain components. Consensus is an important property in blockchain systems which ensures that all the nodes agree on the network state. By a careful consideration of consensus algorithms and protocols, blockchain can attain superior and diverse security features such as data integrity, non-repudiation, and auditability \cite{zheng2017overview,hassan2019blockchain}. Appropriate selection of communication network can have an impact on the decentralization and scalability of the system. For example, if latency is not an issue but decentralization and scalability are required, Proof-of-Work (PoW) can be used. If the system is required to converge in a very short time, 6G can be used with communication-intensive mechanisms like Practical Byzantine Fault Tolerance (PBFT). 

\indent We divide 6G application requirements into two broad categories with the objective of making the blockchain and 6G combination easier to understand. In the first category, called Requirement Group I \mbox{(RG-I)}, we include the performance-related needs on data rates, latency, reliability and massive connectivity. These performance requirements will help enable ubiquitous communication. In the second category, called Requirement Group II \mbox{(RG-II)}, we include the security-related needs on data integrity, non-repudiability, and auditability. The major contributions of this paper are as follows:

\begin{itemize}
    \item We identify the requirements for optimal performance of 6G applications. We divide these requirements into two groups based on traditional and security requirements.
    \item We discuss the combination of blockchain and 6G for these application requirements. The decentralization \& trustlessness and the security features of blockchain will cater to both types of application requirements.
    \item We consider the blockchain employment in an LS-CAS scenario. We derive the time required to detect malicious miners in a blockchain system. By simulation results, we show that blockchain will help detect malicious miners and 6G will help accelerate this detection.
\end{itemize}

\section{6G applications and their requirements}
\begin{figure}[htb]
\centering
	\includegraphics[scale=0.45]{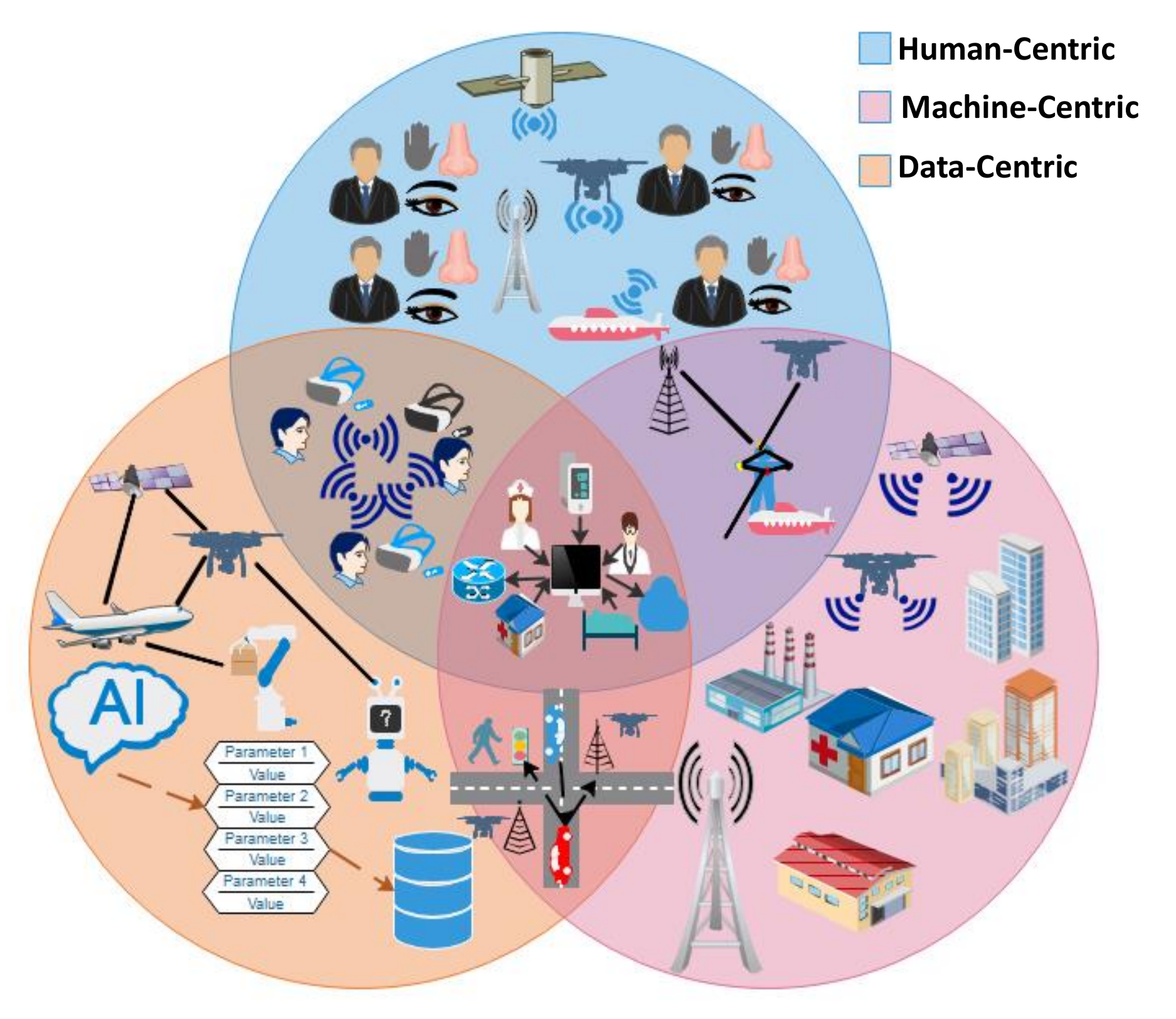}
\caption{6G Applications.}
\label{6G:apps}
\end{figure}

In this section, we discuss some futuristic 6G applications as shown in Figure \ref{6G:apps} and discuss their requirements. 

\subsection{6G Applications}

\subsubsection{Human Bond Communication} This application is concerned with data from all five human senses to allow more expressive, realistic and holistic information exchange between humans and machines. This application would require strict security guarantees because a lot of intimate data would be transmitted.  

\subsubsection{Multi-sensory eXtended Reality Applications} By combining information from human senses, human gestures, surrounding environment, and multiple data sources, XR applications can provide fully-immersive user experiences. Data integrity is required for this application because any data attack by a malicious entity could change the entire user experience.  

\subsubsection{Wearable Technology based Futuristic Applications} Another key area is wearable technology (implantable sensors, wearable clothing, brain-computer-interface (BCI)), which requires ultra-reliability for reliable data exchange. Existing 5G systems fall short in leveraging the numerous potential opportunities beyond traditional healthcare scenarios. 

\subsubsection{Large-scale connected autonomous systems} Another area where 6G can find potential applications is connected robotics and autonomous systems, which include drone-delivery systems, autonomous cars, autonomous drone swarms, vehicle platoons, and autonomous robotics \cite{saad2019vision}.
These applications simultaneously demand all three 5G service classes, and network slicing in 5G may not be the ideal way to achieve the requirements of such applications.    

\subsubsection{Greater Support for Vertical Domains} For vertical industries in which similar products or services are developed, produced, and provided (e.g., manufacturing, energy, health, automation), 3GPP has defined multiple key performance indicators (KPIs) for several core and secondary quality of service (QoS) parameters. 5G massive machine type communication (mMTC) will not be able to cope up with increasing number of connected devices in vertical industries.

\subsection{6G Application Requirements} 
\indent {We divide 6G application requirements into two broad categories with the objective of making it easier to understand blockchain utility}. In the first category, we group those requirements that have always remained a major consideration in all the previous generations of wireless communication systems. These traditional requirements include ultra-reliability, low-latency, enhanced data rates, and massive connectivity. We refer to these as `Requirement-Group-I' \mbox{(RG-I)}. 6G applications will demand several orders of magnitude improvements in \mbox{RG-I} values. In the second category, we include privacy \& confidentiality, data integrity, non-repudiability, and auditability requirements. We refer to these, mostly security-related requirements, as `Requirement-Group-II' \mbox{(RG-II)}. 6G applications are envisioned to use and manipulate large amount of data produced by human senses/organs, and autonomous agents that necessitate the inclusion of both types of requirements as inherent and necessary components/features. 

In 6G vision papers, we find a lot of discussion on various technologies that can help in the further improvements in \mbox{RG-I} values. Some of these front-runner 6G technologies include THz communication, RIS~\cite{saad2019vision, dang2019human, tariq2019speculative} and AI \cite{giordani20196g,letaief2019roadmap}. We anticipate that the advancements in these new technologies and new network architectures will enable ultra-reliable, low-latency, and enhanced broadband connectivity for massive number of devices in 6G communication systems. On the other hand, in 6G literature, there is not much discussion about \mbox{RG-II}. This is due to several reasons. The exact definition and scope of security-related requirements may vary in different application scenarios due to the nature of involved entities such as operators, equipment, and machines, and assigning responsibility for the fulfillment of these requirements is also not straightforward. As the application and use-case landscape in future 6G applications becomes more and more complex, fixing \mbox{RG-II} values also becomes challenging. 

\section{Blockchain and 6G}
In this section, we first discuss blockchain followed by the combination of blockchain and 6G from RG-I and RG-II perspective.

\subsection{Blockchain}
\indent Blockchain is a distributed ledger in which information is stored as a chain of data blocks. Blockchain is an amalgamation of several technologies for network, consensus and automation management. All these technologies have to be carefully combined and selected to attain the desired security features required for the underlying application scenario. As the use cases of blockchain are expanding, so are the number of available options to build a blockchain. With respect to administrative control, blockchain can be either public, consortium or private. Any node can join, leave, read or write on the public blockchain and it is completely decentralized. In consortium and private blockchains, the write access is owned respectively by a group of organizations and a single organization. PoW algorithms provide the greatest amount of security features in terms of data immutability but their use on resource-constrained nodes becomes challenging. PoS variants like dPoS, on the other hand, can be made more secure by engaging large number of verifiers in the network which increases the communication overhead and the time required for reaching consensus. Automation on blockchain is managed through smart contracts which are computer programs stored on the blockchain to define the contractual obligations and enable the automatic transfer of assets between peers when the required conditions are met. 

\subsection{Blockchain and 6G RG-I}
\indent For RG-I targets, 6G is expected to be 3D integrated with infrastructure elements being present in all three dimensions. The management of this infrastructure and asset will be a challenging task. The spectrum, storage and computation sharing models will also become more complex. AI will be an essential part for resource optimization. The management of trained models will become complex. Blockchain will provide the essential trustless environment and security required for the resource and AI management.

\begin{table}[htb]
\centering
\fontsize{6.0pt}{6.0pt}\selectfont
\caption{Blockchain Based Resource and AI model parameter management for RG-I in 6G applications.}
\begin{tabular}{|M{1.2cm} | M{1.2cm} | M{2.8cm} | M{2.0cm} | } \hline 
\rowcolor[HTML]{9B9B9B}
\textbf{Category} & \textbf{Sub-Category} & \textbf{Description} & \textbf{Blockchain Based Solution}  \\ \hline

\multirow{3}{*}{\parbox{1.5cm}{\centering \textbf{Resource Management Solutions)}}}
&  \textbf{Spectrum Management}  & {Spectrum owners can coordinate with each other to provide spectrum resource for high data rates} & {Spectrum usage information can be stored on blockchain} \\ \cline{2-4}

& \textbf{Infrastructure \& Asset Management}  & {3D communication infrastructure is mobile, dense and diverse with complex ownership models. Its management is a challenging task for one entity} &{Infrastructure location, ownership information, usage information, maintenance requirements, and useful life data can be stored on blockchain} \\ \cline{2-4}                

&  \textbf{Computing Power \& Data Storage Management} & {Un-utilized computing power or storage space anywhere in the network can shared to reduce battery drainage, decrease task latency, balance resources, and improve performance} & {Computing power and data space shared information can be stored on blockchain}  

\\ \hline \hline

\textbf{AI Model Parameter Management} & \textbf{AI} & {AI models can be trained for complex operational and environmental optimization tasks} & {Hard trained AI model parameters are securely stored on and retrieved from blockchain} 
  
\\ \hline

\end{tabular}
\label{tab:sols}
\end{table}

\subsubsection{Resource Management Solutions} 
6G applications would demand a large amount of spectrum, computing power, and other available resources and infrastructure. 

\noindent \textbf{Spectrum Management:} {High data rate requirements of 6G applications can be aided by spectrum sharing as data rate is directly proportional to the available bandwidth.} To maximize spectrum utilization, licensed spectrum owners as well as unlicensed spectrum operators in any band can coordinate and cooperate with each other under different terms and conditions automated through smart contracts which are deployed on a blockchain. {Spectrum sharing framework described in \cite{gorla2020blockchain} for 5G can be applied. In this framework, when a user requests the desired bandwidth, the primary operator (operator which has the registration information of the user) checks to see if it has the desired resources. If not, it requests the secondary operator regarding the resource availability. Once confirmed, the primary operator sends the user's information to secondary operator, which sends a service level agreement (SLA) to the primary operator and provides the required permission to use the spectrum. The authorized node verifies the transaction and adds it to the blockchain. This framework can be made more secure for 6G by choosing a consortium blockchain along with appropriate consensus algorithms. 
In such an improved framework, a transaction containing both the operator identities, user identity, and start and end time of spectrum usage is added to current block which is verified by the network. Once verified, the new block containing the spectrum sharing transaction is added to the blockchain.} 

\noindent \textbf{Infrastructure \& Asset Management:} Dense deployment of communication devices in all the three spatial dimensions owned by multiple operators or some specialized asset service providers (SASPs) (e.g., specializing in communication drones, HAPs, submarines) in 6G would be critical for RG-I targets. At the same time, maximum utilization of all the available resources would ensure the quality of service (QoS) of 6G users and would also maximize the revenues of network operators and SASPs. We explain the utility of blockchain through an example of a user wanting to decrease the communication latency by finding the best communication relays provided by the SASPs. In a blockchain-assisted infrastructure and asset management system, an authenticated and registered user will search for the nearest communication relays. The relays check the registration information of the user and its network from the blockchain and then provide the desired connectivity to the user according to the SLAs available in the smart contract for that network. The transaction is recorded on the blockchain after verification through an appropriate consensus algorithm.

\noindent \textbf{Computing Power \& Data Storage Management:} Many 6G applications would require a large amount of data from a very large number of sensors and nodes. For example, to provide fully immersive XR experiences, thousands (if not millions) of extremely small sensors may be required. Processing all this data into meaningful rich information will require a huge amount of computing power. However, even with future enhancements in battery technologies, such intensive applications will severely deplete the battery and storage of mobile devices and computing and storage resources might also be inadequate. For compute power and storage management, authenticated users could use a public blockchain. We can assume a double auction market model where some users in need of computing power or storage space would submit their asks (required resources and price) while others with spare computing power or storage space can place their bids (available resources and price). In every such market round, bids and asks are matched and market clearing price is determined. The double auction algorithm is automated through a smart contract. The transactions are added to new blocks which are verified through consensus and added to the blockchain. 

\subsubsection{AI Model Parameter Management Solutions}
Operational and environmental intelligence may be achieved in 6G networks with the help of AI. With network densification, novel RIS-based channel models, multiple conflicting objectives, and extremely large number of variables, optimization problems in 6G networks would become NP-hard. Instead of applying traditional optimization, deep learning techniques will mostly be used for efficient optimization of network resources in rapidly changing operational and environmental conditions. AI models are difficult to train but very efficient to use and produce results in no time. In this context, blockchain may be used to safely store the hard-found AI model training parameters. 

\subsection{Blockchain and 6G RG-II }
In the following, we provide definitions of data integrity, non-repudiation and auditability. These are essential security features for 6G applications and have not been clearly defined previously in that context.

\noindent \textbf{Data Integrity:} Data integrity refers to the detection of unauthorized changes in data. Data integrity attacks deliberately modify the original information to corrupt communication system for some malicious gains. Data integrity breaches may create safety issues in several control applications in vertical domains and LS-CAS applications. 

\noindent \textbf{Non-repudiation:} Non-repudiation refers to the availability of irrefutable proof of who performed a certain action even if the nodes in the network are not cooperating. As AI is becoming commonplace, we anticipate a very large number of machine-type nodes in 6G applications to mimic some form of human intelligence. In this context, \mbox{non-repudiation} will become an important requirement in several 6G applications.  

\noindent \textbf{Auditability:} Auditability is concerned with the ability to reconstruct complete history of certain event or action from the historical records. In many LS-CAS applications where critical decision making is involved, auditability would be required to fix liability in case of malfunctions, conflicts, or to safeguard commercial and financial interests. 

With these definitions and in order to better understand the advantages provided by blockchain for \mbox{RG-II} targets of 6G applications, we present a brief discussion of security related options available in 4G and 5G systems. The legacy authentication mechanisms in previous generations of communication systems mostly employ symmetric-key cryptography where the same key is used for encryption and decryption of data \cite{schneider2015towards}. Up to 4G communication systems, Authenticated key agreement protocol (AKA) and Extensible Authentication Protocols (EAP) frameworks are largely used. AKA is a challenge-and-response based authentication protocol, while in EAP, the user provides an identity to the eNodeB which is then authenticated by the authentication server. On the other hand, in 5G communication systems, asymmetric public-key-infrastructure (PKI) based cryptography, which provides stronger security properties than symmetric key cryptography, is used \cite{cao2019survey}. There is no protection in 4G communication systems for user data integrity. In 5G communication systems, protection of user data integrity is mandatory over the air interface. Integrity protection is resource demanding, therefore, the maximum data rate for integrity protected data traffic in 5G is limited to 64kbps. 4G communication systems have no provision for \mbox{non-repudiation} because of symmetric key cryptography, while \mbox{non-repudiation} protection is provided by 5G due to PKI based cryptography. In both 4G and 5G communication systems, there is no defined mechanism for auditability of data.

Integration of blockchain in 6G would not only help but also control \mbox{RG-II} targets. Through appropriate selection of network, consensus, and automation management algorithms, blockchain can provide desired levels of data integrity, non-repudiation and auditability. Blockchain allows asymmetric PKI based cryptography and the inclusion of privacy preservation frameworks for greater data \textbf{privacy \& confidentiality}. Blockchain accepts new blocks only after verification through a consensus mechanism among multiple P2P nodes. Every block is linked to its parent block (previous block in the chain) by a cryptographic hash function. This allows \textbf{auditability} and makes it possible to verify data all the way back to the genesis block. \textbf{Data integrity} in any block can be easily verified simply by checking the hash-trees. Moreover, as the blockchain size increases, data tampering becomes even more difficult because of the linkage between all the chained blocks.

\indent In addition to that, some state-of-the-art security technologies in practical communication systems are used in 5G. The encryption system used in 5G is 128-NIA1 which provides a 128 bit security level, i.e., equivalent to that of AES 128. Blockchain utilizes two different levels of security where, for data verification, the data is encrypted. For data storage, the hash of the block is cascaded into the next block. For privacy, 5G uses the Elliptic Curve Integrated Encryption System (ECIES). Here the International Mobile Subscriber Identity (IMSI) of the user is encrypted multiple times to generate different identities every time. In blockchain, privacy is ensured by generating a different key pair for every transaction to avoid linking of transactions. The control plane of 5G is logically centralized using SDN and NFV. This highlights an obvious vulnerability related to availability. Blockchain, being a decentralized network, provides better availability.

\section{Case Study and Simulation Results}
\indent In this section, we present a case study to show how blockchain and 6G combined can provide a fast and secure communication system. We consider an LS-CAS example which is a data- and machine-centric application and large amount of critical data is automatically generated and shared between autonomous nodes. Such applications are already been discussed in 4G and 5G communication systems. However, we will demonstrate that combining blockchain and 4G or blockchain and 5G cannot achieve the same level of synergy that we can obtain with blockahin and 6G. This is because of the fact that superior security features of blockchain require resource-intensive consensus algorithms and superior communication networks. Therefore, when 6G speeds are combined with blockhain security we achieve the desired goal of truly fast and secure communication.   

\subsection{LS-CAS Scenario}
Our LS-CAS scenario consists of autonomous vehicles \& delivery drones (collectively referred to as User Equipments (UE)) and Road Side Units (RSU). We assume that some RSUs are fixed while others are drone-mounted. In this application, we have U2U (user equipment to user equipment), U2I (user equipment to infrastructure) and I2I (infrastructure to infrastructure) communications. UEs and RSUs together form a large-scale wireless-connected distributed autonomous system. We assume UEs are equipped with several sensors and state-of-the art camera systems. The data generated by UEs might represent real-time road maps, location information, infotainment, RSU reputation information, sensor readings, or any other information related to UE safety, transportation or entertainment needs. This data should be shared in the network with low latency while ensuring data integrity which is important for safe navigation and other reasons. In a scenario where this system is under attack from some malicious actors (RSUs and colluding vehicles) inside the network who can tamper data for their advantage, we need a mechanism to detect data tampering and also to recognize malicious actors. With the introduction of blockchain and its features, such data integrity attacks and bad actors can be easily recognized.

\subsection{Secure Enhanced dPoS Algorithm for LS-CAS}
\begin{figure}[htb]
\centering
	\includegraphics[scale=0.35]{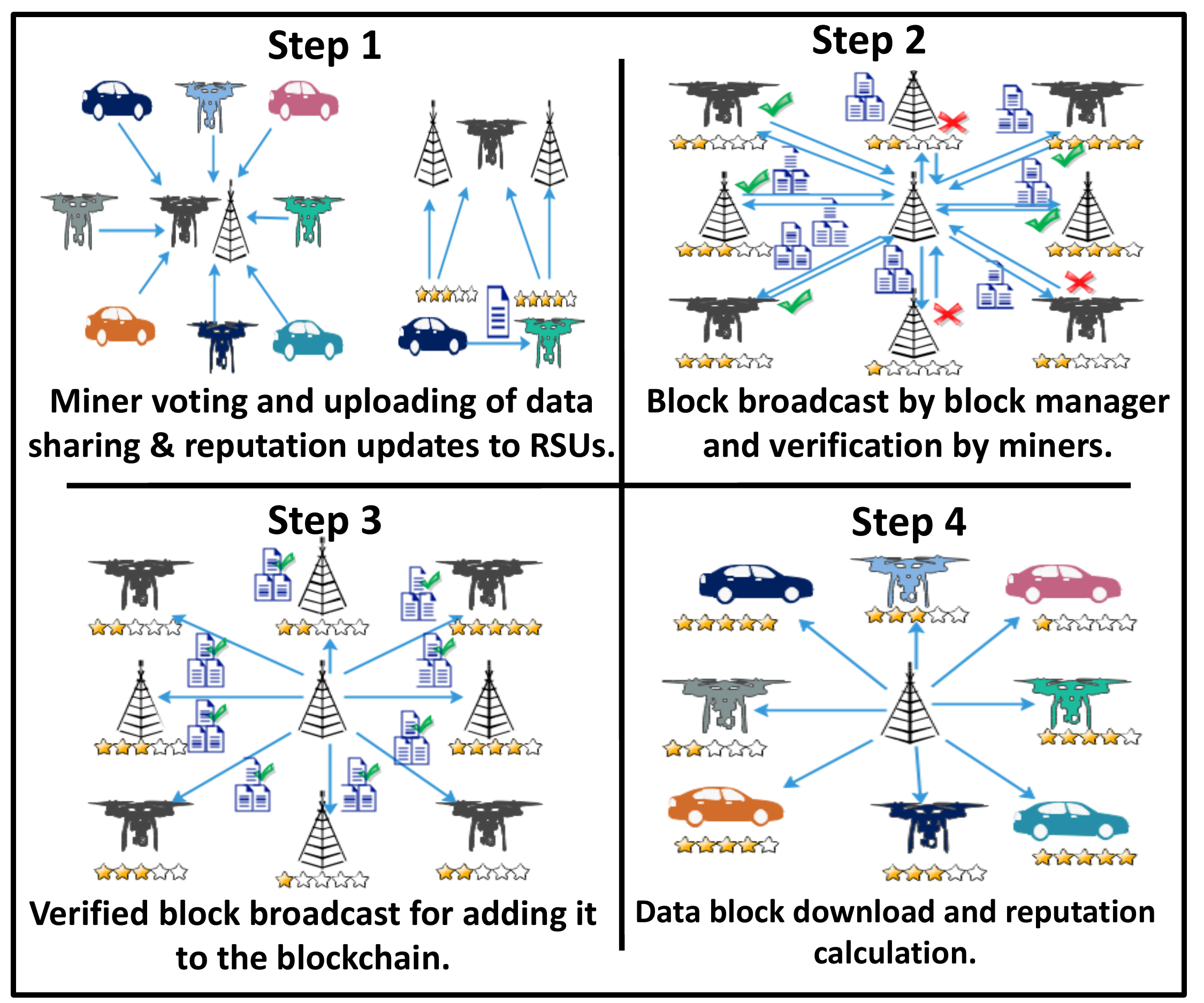}
\caption{Blockchain-based reputation and dPoS Implementation for LS-CAS Application.}
\label{fig:scenario}
\end{figure}

We consider a blockchain-based setup similar to \cite{kang2019toward}. This blockchain uses a secure and enahnced dPoS alogrithm and there are numerous safegaurds for the protection of shared data. We assume RSUs have the necessary resources to implement and store the blockchain. The data shared among the UEs is sent to the RSUs, which run a dPoS consensus algorithm for block mining. We assume that RSUs are not fully trusted and can get compromised. Additionally, some UEs can also collude with the compromised RSUs. Therefore, miner reputations are updated after a complete round of data exchange and the record is uploaded to the blockchain. In the following, we explain one round of block creation and reputation updates. This process is also depicted in Figure~\ref{fig:scenario}.

\begin{enumerate}
\item {First, the stakeholders (vehicles/drones) participate in a voting process to determine active and standby miners according to the reputation scores. Active miners are a pre-defined number of higher reputation miners from all the miners, and they act as block managers in a round-robin fashion for the following rounds. Real-time data is shared among UEs and the data sharing record is shared with the nearest available RSU. UEs also upload recently calculated reputation scores to the nearest RSU. RSUs route this data to the block manager of that round.} 
\item {Active and standby miners are split into different types based on their reputation scores. The block manager designs a smart contract for each type and broadcast the unverified data block along with the different smart contracts. Smart contracts are designed such that a verifier gets maximum utility only if it attempts the respective smart contract. The block is verified and the verification results are audited by the local neighborhood after which, the block is sent back to the block manager.}
\item Block manager receives the verification reports and creates a new data block based on 2/3 majority consensus basis. After consensus, the block manager broadcasts new block and the RSUs add this block to their local blockchain copies.  
\item UEs download the latest data block from their nearest RSU, check the accuracy of their previous transactions, and accordingly update the reputation score of the RSU for the next round.   
\end{enumerate}
There are numerous safeguards in this dPoS scheme to allow for the detection of malicious actors and collusion attacks. However, it is also obvious that the detection of any malicious activity largely depends on the amount of time required to complete different tasks in each round. We can divide the latency of different steps in each round into transmission latency, computational latency
and information diffusion latency. Due to the availability of relatively fast processors in vehicles and RSUs, we neglect the computational latency and thus the time for one round of this step will depend on the network speeds and network scale. 
  
\subsection{The role of Communication Network}
\indent To show the role of communication networks (4G/5G/6G) we perform some simulations. In these simulations, we assume a total area of 150km$^2$. UE positions are randomly initialized and RSUs are uniformly deployed across the network. The positions and the range of RSUs are set according to the density of network deployment. {The weight of positive and negative interactions is set as 0.4 and 0.6 respectively. The probability of successful message transmission is 0.7. These parameters are derived from \cite{kang2019toward}. The adjustment factor for number of hops is set to be 0.75.} The reputation scores are computed using multi-weight subjective logic (MWSL) model \cite{kang2019toward}. We consider four different network scales, i.e., small-scale, medium scale, large-scale, and very-large-scale. Different parameters required in the simulation (e.g., average number of hops to block manager, types of verifiers, number of RSUs with UE records) are adjusted according to the network scale. In these simulations, we consider an attack scenario where a miner starts to behave maliciously after 20 rounds. The malicious miner also colludes with 25\%, 33\% and 50\% UEs in order to get high reputation scores. We consider the ability of the blockchain-based scheme to detect the malicious miner in 4G, 5G and 6G networks. Important simulation parameters are given in Table~\ref{tab:simparas}.

\begin{figure}[htb]
\centering
	\includegraphics[scale=0.3]{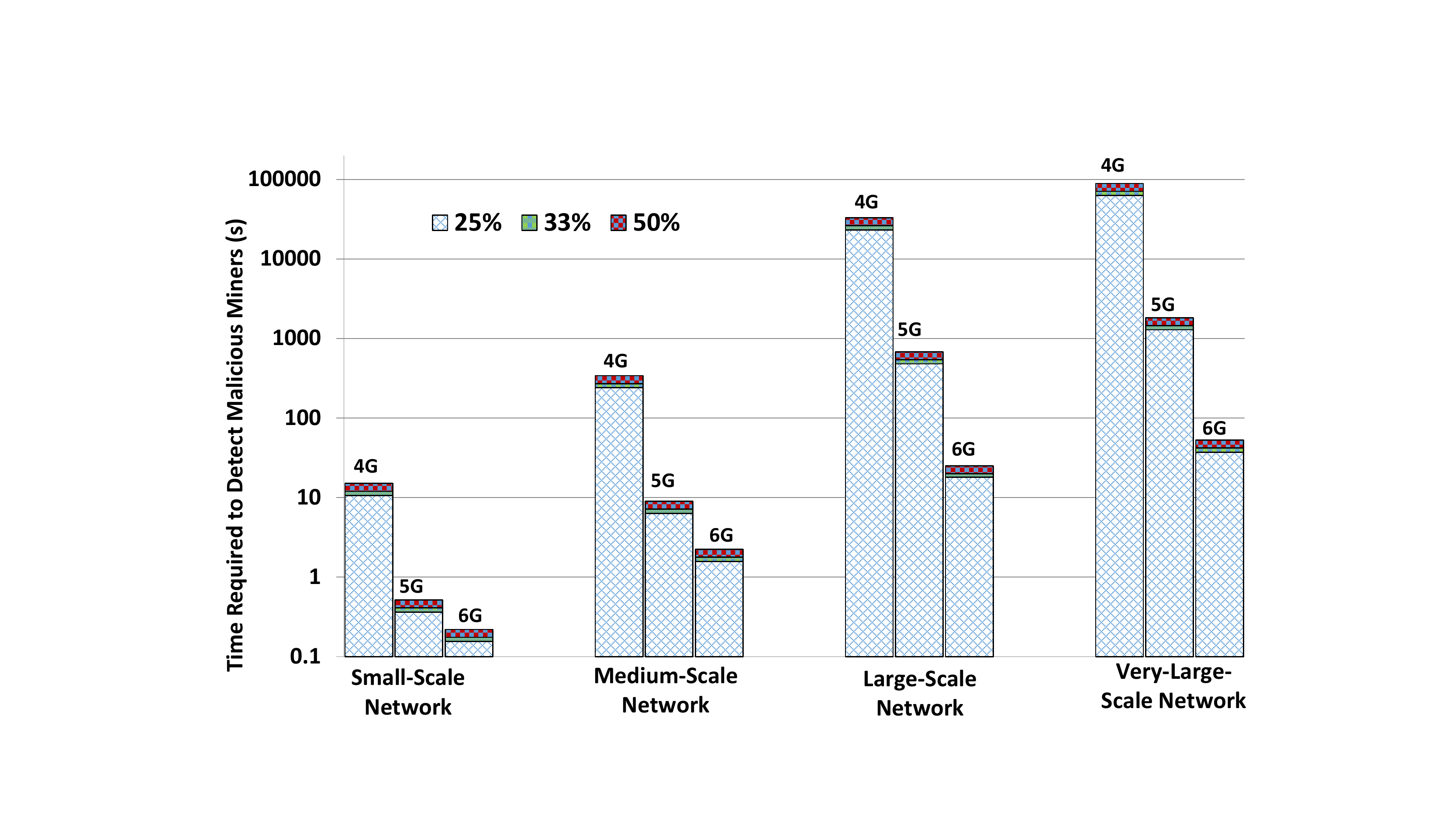}
\caption{Time Required to Detect Malicious Miner for Different Collusion Rates in Different Sized Networks.}
\label{fig:res}
\end{figure}

In Figure~\ref{fig:res}, we plot the amount of time required to detect malicious miner for different network sizes in 4G/5G/6G for different attack scenarios. As we increase the network size, the amount of time required for the detection of malicious miners increases. Similarly, for the same network size, as we increase the percentage of colluding UEs, the amount of time required to detect a malicious miner also increases. The performance of 4G network is only adequate in small and medium-scale network where at 50\% collusion it is able to detect the malicious miner respectively in 15s and 340s respectively. In large and very-large-scale networks, at 50\% collusion, 5G network requires 681s and 1826s respectively to detect a malicious miner. On the other hand, 6G network only requires 25s and 53s respectively in large and very-large-scale networks to detect a malicious miner at 50\% collusion.

\begin{table}[htb]
\centering
\fontsize{6.0pt}{6.0pt}\selectfont
\caption{Parameters and their values}
\begin{tabular}{| M{2.0cm} | M{1.2cm} | M{1.2cm} | M{1.2cm} |M{1.2cm} | } \hline 
\rowcolor[HTML]{9B9B9B}
\textbf{Parameter} & \textbf{Small-Scale Network} & \textbf{Medium-Scale Network} & \textbf{Large-Scale Network} & \textbf{Very-Large-Scale Network} \\ \hline
Total number of active and standby miners  & 100 & 1000 & 10000 & 20000 \\ \hline
Total number of vehicular and drone users & 100 & 1000 & 10000 & 20000\\ \hline
Vote Size  & 1KB & 10KB &100KB & 200KB\\ \hline
UEs and RSUs download and upload speeds  & 10Mbps(4G), 500Mbps(5G), 100Gbps(6G) & 10Mbps(4G), 500Mbps(5G), 100Gbps(6G) & 10Mbps(4G), 500Mbps(5G), 100Gbps(6G) & 10Mbps(4G), 500Mbps(5G), 100Gbps(6G) \\ \hline
Data block size before verification  & 10KB & 100KB & 5MB & 10MB \\ \hline
Reputation block size before verification  & 1.5KB & 15KB & 150KB & 300KB \\ \hline
Size of smart contract  & 2KB & 15KB & 150KB& 200KB\\ \hline
Types of Verifiers  & 10& 10& 10& 10\\ \hline
Number of active miners  & 15 & 41 & 199 & 255\\ \hline
Number of RSUs with UE data record  &  [10, 40] &  [100, 400]&  [1000, 4000]&  [1500, 6000] \\ \hline
Maximum end to end number of hops  & 8 & 23 & 71 & 100\\ \hline

\end{tabular}
\label{tab:simparas}
\end{table}

\begin{figure}[htb]
\centering
	\includegraphics[scale=0.55]{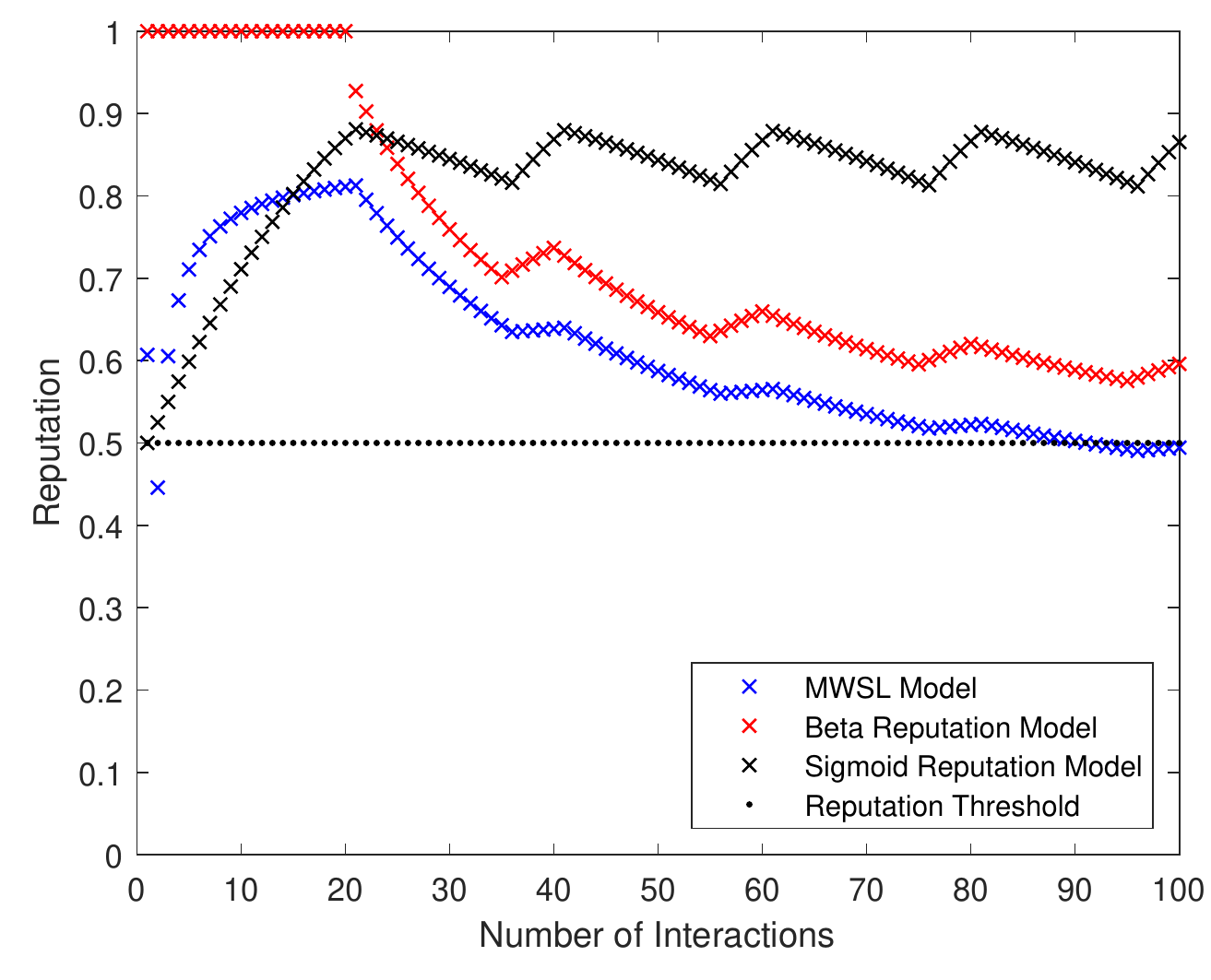}
\caption{Update sensitivity of different reputation schemes.}
\label{fig:rep}
\end{figure}

\indent In order to clarify the importance of blockchain in this scenario, we consider an adversary, which behaves honestly for 20 interactions and then switches between malicious and honest behavior for 15 and 5 interactions alternatively. We use the blockchain with MWSL model as well as blockchain with beta and sigmoid reputation models. In beta reputation model, beta probability density function is used to combine feedback and derive reputation. In sigmoid model, reputation is calculated as a sigmoid function of an overall impact of honest and malicious behavior. Using appropriate parameters and considering 33\% network collusion, we observe that some blockchains are able to detect malicious miners, while others are not as shown in Figure~\ref{fig:rep}. Even using 6G will not help in those cases. Using the latency results of 6G, we will see that malicious miners are detected the earliest when used 6G with an appropriate blockchain model. This result shows that in LS-CAS, appropriate selection of blockchain structure is necessary for detecting malicious activity and therefore, improve the system integrity. Along with blockchain, 6G is the most appropriate technology that will facilitate timely detection. In that sense, blockchain and 6G will form an ideal combination for the application used i.e., LS-CAS.

\indent {These simulation results are very promising and suggest the use of more secure blockchain implementations in 6G are possible as both complement each other. Secure consensus algorithms enhance security of 6G applications while 6G enables their implementation through its faster speeds. At the same time, the creation of a trustless environment in 6G by more secure blockchain implementations would benefit RG-I by eliminating under-utilization of critical resources deployed under complex ownership and sharing models. Some challenges for wider blockchain implementations in 6G would require further research in the following directions:}

\begin{itemize}
  \item {Sharding and sub-blockchain techniques could be utilized for further reduction in convergence times in very large blockchain networks. }
  \item {Smart contracts optimization techniques are necessary for decreasing block size and consensus latency. Smart contracts should also be written extremely carefully to make them less vulnerable to hackers.}
  \item {Larger network sizes translate to larger storage requirements. Off-chain storage can be used and a signature associated with the block can be stored on the chain.}	
    \item {Without compromising the security features, there is a clear need for less resource intensive consensus algorithms.}
	\end{itemize}

\section{Conclusion}
In this paper, we have discussed the potential of blockchain and 6G for future communication and highlighted a synergy between them. We have divided 6G application requirements into performance related \mbox{(RG-I)} and security related \mbox{(RG-II)} groups with the objective of making the synergy more understandable. We have shown that the trustless nature of blockchain would make it easier to manage and audit 3D network resources and AI model parameters in 6G networks with complex ownership models. This flexible use of increasingly large and complex network resources in 6G with the help of blockchain would significantly facilitate RG-I targets. Furthermore, through appropriate selection of blockchain type and consensus algorithms, RG-II needs of 6G applications could also be readily addressed. Therefore, blockchain and 6G combined can provide secure and ubiquitous communication.

\bibliographystyle{IEEEtran}   
\bibliography{main} 

\end{document}